\documentclass[epjST]{svjour}
\usepackage{graphics}
\usepackage{graphicx}
\usepackage{dcolumn}
\usepackage{amsmath}
\usepackage{amssymb}
\usepackage{color}

\newif\ifgraph

\graphtrue
\begin{document}
\title{Active Brownian motion in a narrow channel}
\author{Xue Ao\inst{1}, Pulak~K.~Ghosh\inst{2},
Yunyun Li\inst{3}\fnmsep\thanks{\email{yunyunli@tongji.edu.cn}}, Gerhard Schmid\inst{1},
Peter H\"anggi\inst{1,3} \and Fabio~Marchesoni\inst{3,4}}
\institute{Institut f\"ur Physik, Universit\"at Augsburg, D-86135
Augsburg, Germany \and Department of Chemistry, Presidency
University, Kolkata - 700073, India \and Center for Phononics and
Thermal Energy Science, School of Physic Science and Engineering,
Tongji University, Shanghai 200092, People's Republic of China \and
Dipartimento di Fisica, Universit\`{a} di Camerino, I-62032 Camerino,
Italy}
\abstract{
We review recent advances in rectification control of artificial
microswimmers, also known as Janus particles, diffusing along narrow,
periodically corrugated channels. The swimmer self-propulsion
mechanism is modeled so as to incorporate a nonzero torque
(propulsion chirality). We first summarize the effects of chirality
on the autonomous current of microswimmers freely diffusing in
channels of different geometries. In particular, left-right and
upside-down asymmetric channels are shown to exhibit different
transport properties. We then report new results on the dependence of
the diffusivity of chiral microswimmers on the channel geometry and
their own self-propulsion mechanism. The self-propulsion torque turns
out to play a key role as a transport control parameter.
} 
\maketitle
\section{Introduction}
\label{intro}
Rectification of Brownian motion in a narrow, periodically corrugated
channel has been the focus of a concerted
effort~\cite{ChemPhysChem,RMP2009,Denisov} aimed at establishing net
particle transport in the absence of external biases. To this purpose
two basic ingredients are required: a spatial asymmetry of the
channel and a time correlation of the perturbations, random or
deterministic, applied to the diffusing particles. The ensuing
spontaneous current is a manifestation of the so-called ratchet
effect~\cite{RMP2009}. Typically, demonstrations of the ratchet
effect had recourse to external {\it unbiased} time-dependent drives
(rocked and pulsated ratchets). Rectification induced by
time-correlated, or colored, non-equilibrium fluctuations (thermal
ratchets) is conceptually feasible, but has been, so far, of limited
practical use. The idea in itself, however, is appealing: The
diffusing particles would harvest kinetic energy directly from their
non-equilibrium environment, without requiring any externally applied
field at all, and transport would ensue as an {\it autonomous}
symmetry-directed particle flow.

\begin{figure}      
\centering
\includegraphics[width=0.95\linewidth]{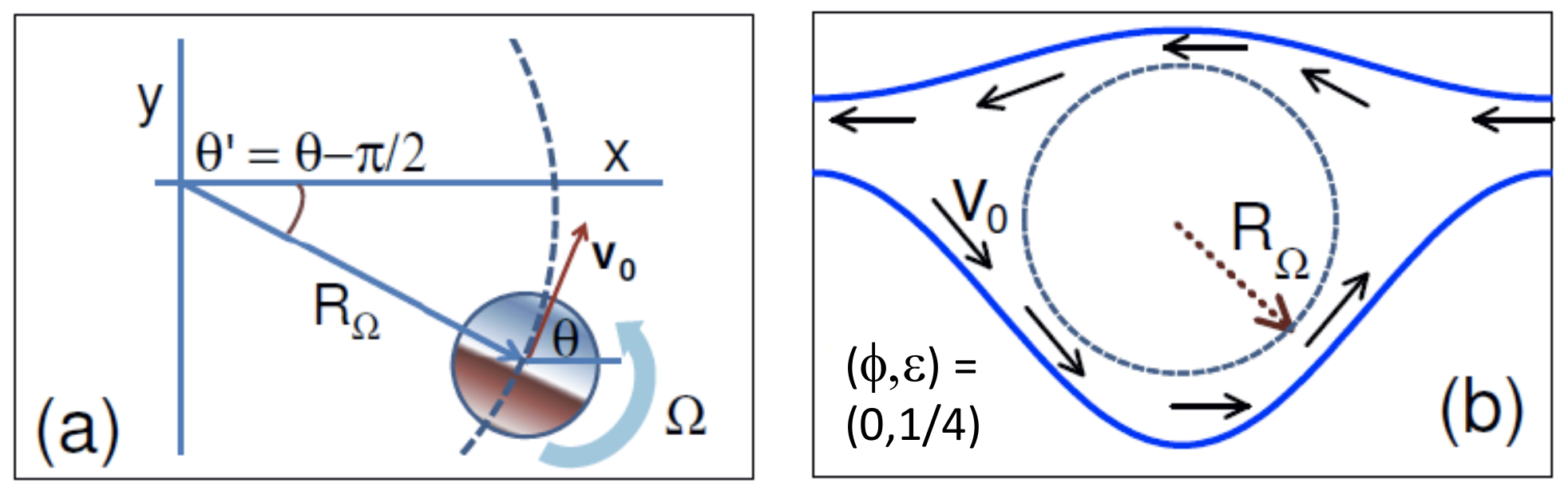}
\caption{(Color online) Levogyre Janus particle in a narrow channel:
(a) noiseless particle with velocity ${\vec v}_0$ and torque
frequency $\Omega>0$, Eq. (\ref{LE}), moving along a circular arc of
radius $R_\Omega$ (dashed line); (b) upside-down asymmetric channel
compartment, Eq. (\ref{wx}) with $\phi=0$ and $\epsilon=0.25$. In (b)
the open boundary-trajectories and a closed circular trajectory of
radius $R_\Omega$ are drawn for explanatory purposes (see Sec.
\ref{upsidedown} for actual simulation data).} \label{F1}
\end{figure}

To enhance rectification of time correlated diffusion in a modulated
channel with zero drives, we recently proposed~\cite{MSshort} to make
use of a special type of diffusive tracers, namely of active, or
self-propelled, Brownian particles. Self-propulsion is the ability of
most living organisms to move, in the absence of external drives,
thanks to an ``engine'' of their own~\cite{Purcell}. Optimizing
self-propulsion of micro- and nano-particles (artificial
microswimmers) is a growing topic of today's
nanotechnology~\cite{Schweitzer,Rama,Ebeling}. Recently, a new type
of artificial microswimmers has been synthesized \cite{Granick,Chen},
where self-propulsion takes advantage of the local gradients
asymmetric particles can generate in the presence of an external
energy source (self-phoretic effects). Such particles, called Janus
particles (JP), consist of two distinct ``faces", only one of which
is chemically or physically active. Thanks to their functional
asymmetry, these active particles can induce either concentration
gradients (self-diffusiophoresis) by catalyzing a chemical reaction
on their active surface~\cite{Paxton1,Gibbs,Bechinger}, or thermal
gradients (self-thermophoresis), e.g., by inhomogeneous light
absorption \cite{Sano} or magnetic excitation~\cite{ASCNano2013JM}.

A self-propulsion mechanism acts on an pointlike particle by means of
a force and, possibly, a torque. In the absence of a torque, the line
of motion is directed parallel to the self-phoretic force and the JP
propels itself along a straight line, until it changes direction, due
to gradient fluctuations~\cite{Sen_propulsion} or random collisions
against other particles or geometric boundaries~\cite{Vicsek}. This
is the highly stylized case mostly studied in the recent literature,
where, for simplicity, the JPs are assumed to be rotationally
symmetric around their line of motion (symmetric JPs). In the
presence of an additional torque the self-phoretic force and the line
of motion are no longer aligned and the microswimmer tends to execute
circular orbits~\cite{Lowen,Julicher}.

Active chiral motion has long been known in biology
\cite{Brokaw,Julicher,Volpe} and more recently observed in
asymmetrically propelled micro- and nano-rods: A torque can be
intrinsic to the propulsion mechanism, due to the presence of
geometrical asymmetries in the particle fabrication, engineered or
accidental (asymmetric JP's)~\cite{LowenKumm,Ibele,composite}, or
externally applied, for instance, by laser irradiation \cite{Sano} or
hydrodynamic fields~\cite{Stark}. In the finite damping regime, the
Lorentz force exerted by a magnetic field on a charged active
Brownian particle also amounts to an external
torque~\cite{Schimansky}; however, the effects of such magnetic
torques vanishes in the overdamped limit~\cite{Kline}.

In this paper we discuss the interplay of chiral propulsion and
channel spatial asymmetry on controlling autonomous rectification. In
Sec. \ref{model} we remind that active Brownian motion is time
correlated {\it per se}, which means that transport control of, say,
a JP with assigned self-propulsion properties, can only be achieved
by suitably tailoring the channel boundaries. In Sec.
\ref{rectificaton} we briefly review autonomous rectification of JP's
with and without external torque. In both cases the net on-plane
drive exerted on the particle is null (unbiased diffusion). We
distinguish between two categories of channels, left-right asymmetric
channels, Sec. \ref{leftright}, where even nonchiral JP's can be
rectified, and upside-down asymmetric channels, Sec.
\ref{upsidedown}, where particle ratcheting requires a nonzero
torque. Based on numerical evidence, in Sec. \ref{asymmetry} we
establish the minimal asymmetry conditions that make rectification of
active Brownian particles possible. Finally, in Sec. \ref{diffusion}
we present new results on active diffusion in channels of various
geometries. In particular, we show that diffusion of JP's is not
controlled by the channel geometry as much as by the angular
asymmetry of the self-propulsion mechanim. The diffusion of nonchiral
and chiral JP's is investigated in Secs. \ref{diffnonchiral} and
\ref{diffchiral}, respectively. Ideas for future work are discussed
in the concluding Sec. \ref{conclusions}.

\begin{figure}
\centering
\includegraphics[width=0.45\textwidth]{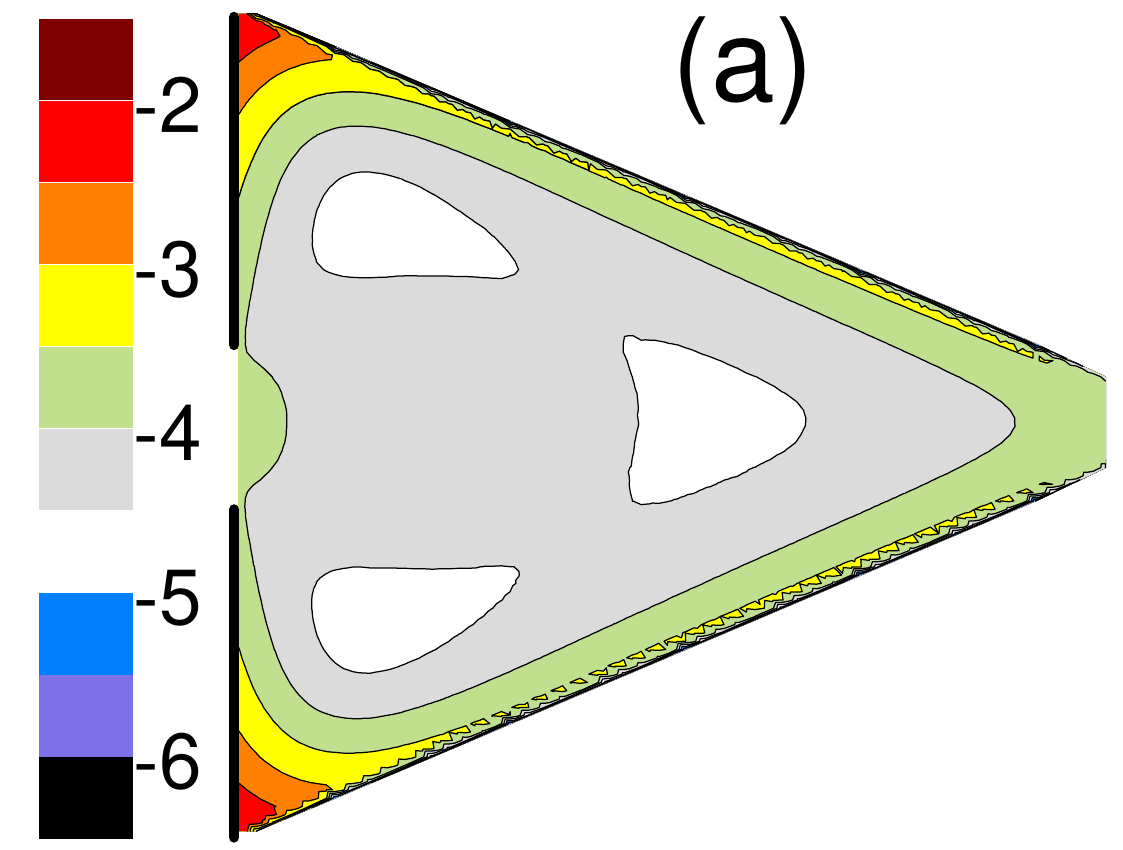}
\includegraphics[width=0.45\textwidth]{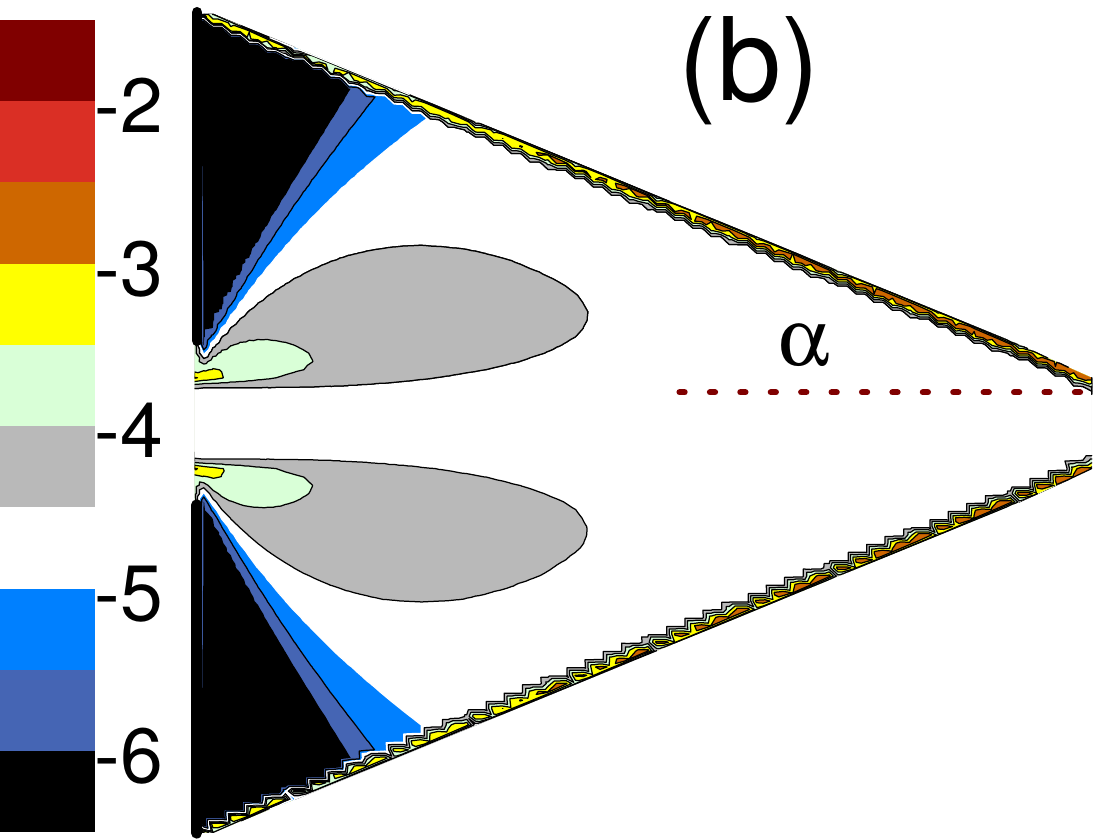}
\caption{(Color online) Logarithmic contour plots of the stationary
particle density, $P(x,y)$, in a triangular compartment of size
$x_L=y_L=1$ and pore width $\Delta=0.1$; the compartment acts as a
funnel of angular width $2\alpha$. Other simulation parameters are:
$D_\theta =0.006$, $v_0=1$ and (a) $D_0=0.03$, (b) $D_0=0$ (no
thermal noise); sliding b.c. have been adopted throughout. Both
densities are singular at the corners and along the side walls, where
graphics resolution effects are apparent. \label{F2}}
\end{figure}

\section{Model} \label{model}

In order to avoid unessential complications, we restrict this report
to the case of 2D channels and pointlike artificial microswimmers of
the JP type \cite{Granick}.
A chiral JP gets a continuous push from the
suspension fluid, which in the overdamped regime amounts to a
rotating self-propulsion velocity ${\vec v_0}$ with constant modulus
$v_0$ and angular velocity $\Omega$. Additionally, the
self-propulsion direction varies randomly with time constant
$\tau_\theta$, under the combined action of thermal noise and
orientational fluctuations intrinsic to the self-propulsion
mechanism. Accordingly, the microswimmer mean free self-propulsion
path approximates a circular arc of radius $R_\Omega=v_0/|\Omega|$
and length $l_\theta=v_0\tau_\theta$ \cite{Lowen}. Chiral effects are
prominent when $R_\Omega \lesssim l_\theta$, or equivalently,
$|\Omega|\tau_{\theta} \gtrsim 1$ (strong chirality regime).

The bulk dynamics of such an overdamped chiral JP obeys the Langevin
equations (LE) \cite{Lowen}
\begin{eqnarray}
\label{LE} \dot x &=& v_0\cos \theta +\xi_x(t) \\ \nonumber \dot y
&=& v_0\sin \theta +\xi_y(t) \\ \nonumber \dot \theta &=&\Omega
+\xi_\theta(t),
\end{eqnarray}
where ${\bf r}=(x,y)$ are the coordinates of a particle subject to
the Gaussian noises $\xi_{i}(t)$, with $\langle \xi_{i}(t)\rangle=0$
and $\langle \xi_{i}(t)\xi_{j}(0)\rangle=2D_0\delta_{ij}\delta (t)$
for $i=x,y$, modeling the equilibrium thermal fluctuations in the
suspension fluid. The channel is directed along the $x$ axis, the
self-propulsion velocity is oriented at an angle $\theta$ with
respect to it and the sign of $\Omega$ is chosen so as to coincide
respectively with the positive (levogyre) and negative (dextrogyre)
chirality of the swimmer, see Fig. \ref{F1}(a). The orientational
fluctuations of the propulsion velocity are modeled by the Gaussian
noise $\xi_\theta(t)$ with $\langle \xi_{\theta}(t)\rangle=0$ and
$\langle \xi_{\theta}(t)\xi_{\theta}(0)\rangle=2D_{\theta}\delta(t)$,
where the noise strength $D_\theta$ is the relaxation rate of the
self-propulsion velocity, $D_{\theta}=2/\tau_{\theta}$, see Sec.
\ref{diffnonchiral} for more details.

In this work the JP's were assumed to be pointlike as we intended to
focus on the causes of autonomous transport and its control. However,
numerical and experimental evidence clearly show that the dynamical
parameters of a real JP in the bulk, i.e., its self-propulsion speed,
friction coefficient, thermal and rotational diffusion coefficients
\cite{finitesize} and effective shape, all depend on its size as well
as on its shape \cite{Spagnolie}. These effects can be accounted for,
at least qualitatively, by an appropriate choice of the free
parameters introduced in our model Eqs. (\ref{LE}).

The simplifications introduced here are not limited to the
dimensionality of the channel or the size of the particle. All noise
sources in Eq. (\ref{LE}) have been treated as independently tunable,
although, strictly speaking, thermal and orientational fluctuations
may be statistically correlated (see, e.g., \cite{Volpe}). Moreover,
we ignored hydrodynamic effects, which not only favor clustering in
dense mixtures of JP's \cite{Ripoll,Buttinoni}, but may even cause
their capture by the channel walls \cite{Takagi}. However, we made
sure that the parameters used in our simulations are experimentally
accessible, as apparent on expressing times in seconds and lengths in
microns (see Refs \cite{Bechinger,Volpe} for a comparison).

When confined to a channel directed along the $x$ axis, the particle
transverse coordinate, $y$, is bounded by the wall functions
$w_{\pm}(x)$, $w_{-}(x)\leq y \leq w_{+}(x)$. All wall geometries
considered below are periodic with compartment length $x_L$, namely
$w_{\pm}(x+x_L)=w_{\pm}(x)$. The channel compartments are connected
by pores of width $\Delta$, much narrower than their maximum
cross-section.

Simulating a constrained JP requires defining its collisional
dynamics at the boundaries. For the translational velocity $\vec{\dot
r}$ we assumed elastic reflection. Regarding the
coordinate $\theta$ we considered two possibilities: \\
{\it (a) frictionless collisions, $\theta$ unchanged.} The active
particle slides along the walls for an average time of the order of
$\tau_\theta$, until the $\theta (t)$ fluctuations redirect it toward
the interior of the compartment. For simplicity, all simulation
results presented in this report have been obtained for sliding b.c.
The panels of Fig. \ref{F2} show how the stationary particle
probability density $P(x,y)$ accumulates along the boundaries; this
effect is even stronger in the
noiseless case, $D_0=0$; \\
{\it (b) rotation induced by a tangential friction, $\theta$
randomized.} These b.c. cause the particle to diffuse away from the
boundary, which, as discussed below, tends to weaken the
rectification effect. Note that, as pointed out in Refs.
\cite{inertia}, should one assume elastic boundary reflection for
both $\vec{\dot r}$ and $\vec{v}_0$, then the self-propelled motion
of a JP would coincide with an ordinary Brownian motion with finite
damping constant, $\gamma=2/\tau_\theta$, and self-diffusion constant
$D_s$ defined in Sec. \ref{diffnonchiral}. Being that an equilibrium
random process, it could not be rectified, no matter what the spacial
asymmetry of the channel.

We further stress that on modeling the boundary conditions we heavily
relied on the pointlike particle assumption to neglect (i) the
dependence of the self-propulsion mechanism on the particle distance
from the walls; (ii) the hydrodynamic interactions between particle
and walls \cite{Spagnolie}; and (iii) the ensuing particle
accumulation against the walls \cite{Takagi}.

Finally, Eqs. (\ref{LE}) have been numerically integrated by using a
standard Milstein algorithm \cite{MSshort} with a very short time
step, $10^{-5} - 10^{-7}$, to ensure numerical stability. As initial
conditions we have assumed that at $t=0$ the particle is uniformly
distributed with random orientation in a channel compartment located
between $x=0$ to $x= x_L$. The total observation time was set to
$10^4 \times \tau_\theta$, or $10^4 \times \Omega^{-1}$, or $10^4$,
whichever is greater, so that effects due to the initial conditions
and transient processes can be neglected. The results reported in the
figure shown here have been obtained by ensemble averaging over $10^4
- 10^6$ trajectories, depending on the observable.

\begin{figure}[bp]
\centering
\includegraphics[width=0.45\textwidth]{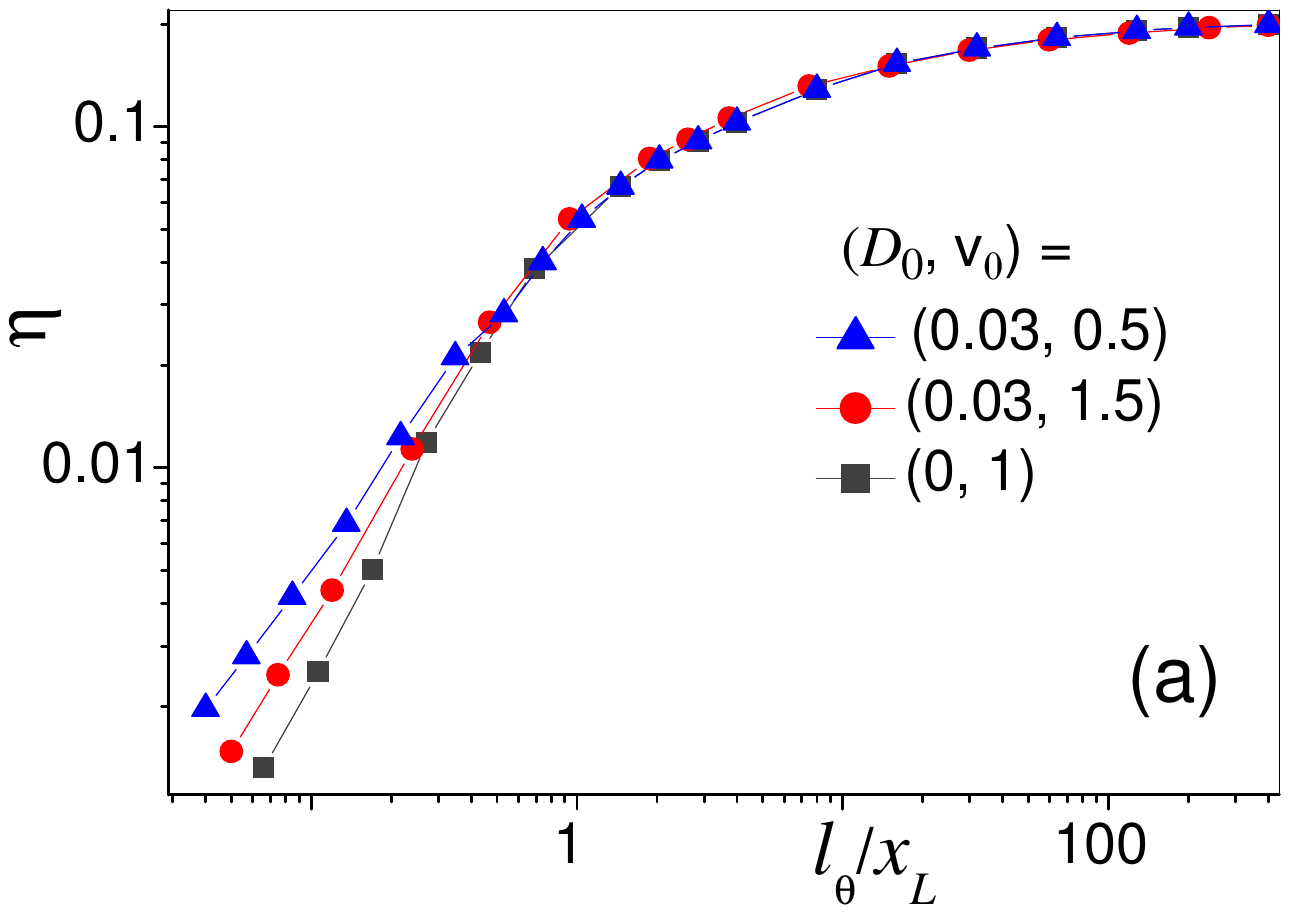}
\includegraphics[width=0.45\textwidth]{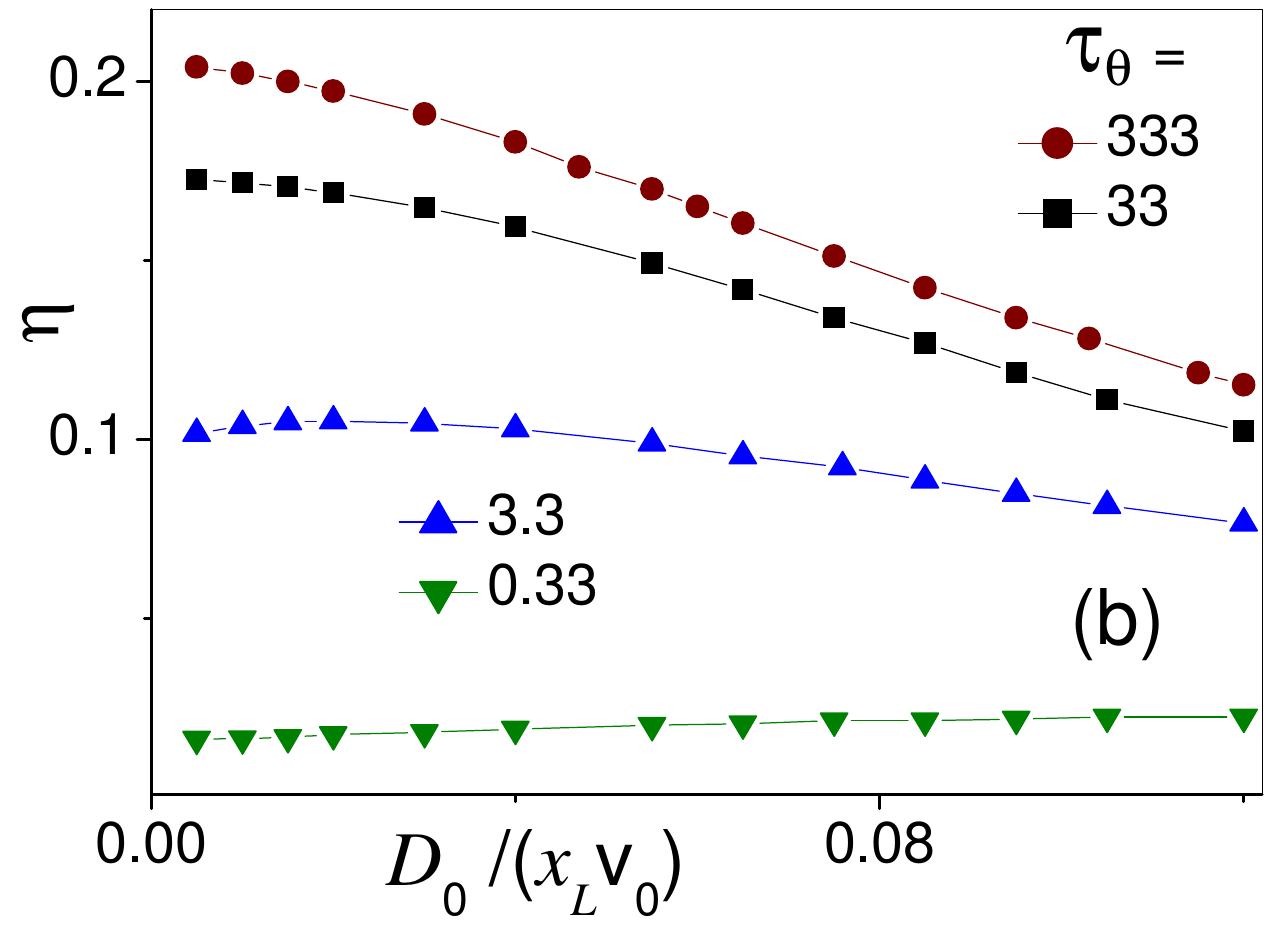}
\caption{(Color online) Rectification of a nonchiral JP in a
triangular channel with compartment geometry as in Fig. \ref{F2}. (a)
rectification power, $\eta$ (solid symbols) vs $l_\theta=v_0
\tau_\theta$ for different $D_0$ and $v_0$; (b) $\eta$ vs. $D_0$ for
$v_0=1$ and different $\tau_\theta$. The particle flow is oriented to
the right, i.e., $\bar v>0$. \label{F3}}
\end{figure}

\section{Autonomous currents} \label{rectificaton}

We discuss first autonomous rectification of a JP in two different
classes of 2D asymmetric channels. To characterize its drift  in the
absence of external biases, we introduce the {\it rectification
power}
\begin{equation}
\label{RP} \eta={|\bar v|}/v_0,
\end{equation}
where the net drift velocity of the particle, $\bar v = \lim_{t\to
\infty}\langle x(t)-x(0) \rangle/t$, is expressed in units of its
self-propulsion velocity, $v_0$.

\subsection{Left-right asymmetric channels} \label{leftright}

\begin{figure}[bp]
\centering
\includegraphics[width=0.65\textwidth]{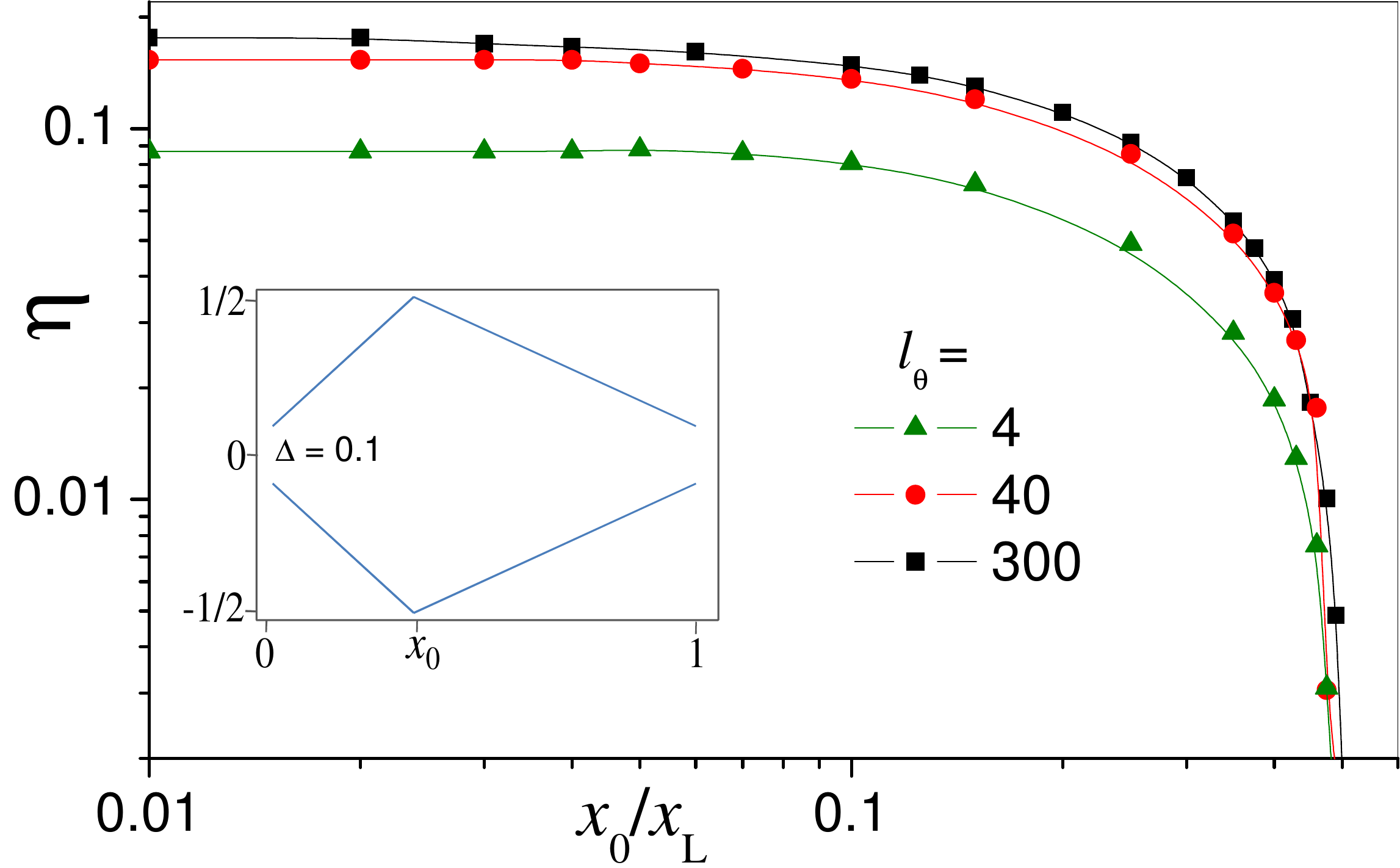}
\caption{(Color online) Rectification of nonchiral JP with $v_0=1$ in
asymmetric channels with different geometries. A typical compartment
is sketched in the inset: $x_L$, $y_L$, and $\Delta$ are as in Fig.
\ref{F2}, but the corners are shifted by $x_0$. Rectification power
$\eta$ vs. $x_0$ for $D_0=0.03$, and different
$D_\theta=2v_0/l_\theta$. The particle orientation is as in Fig.
\ref{F3} for all $x_0=0$. \label{F4}}
\end{figure}

\begin{figure}
\centering
\includegraphics[width=0.45\textwidth]{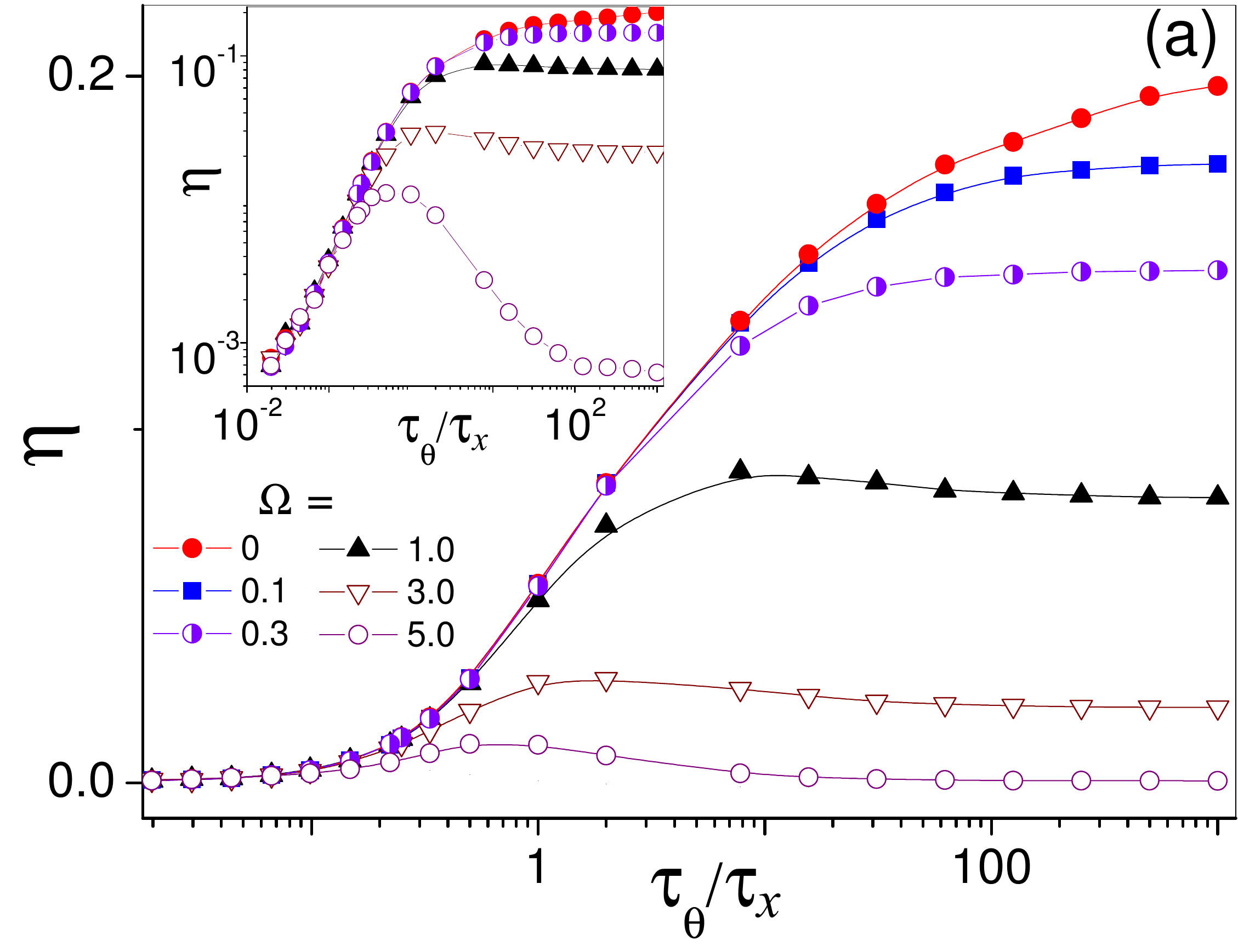}
\includegraphics[width=0.45\textwidth]{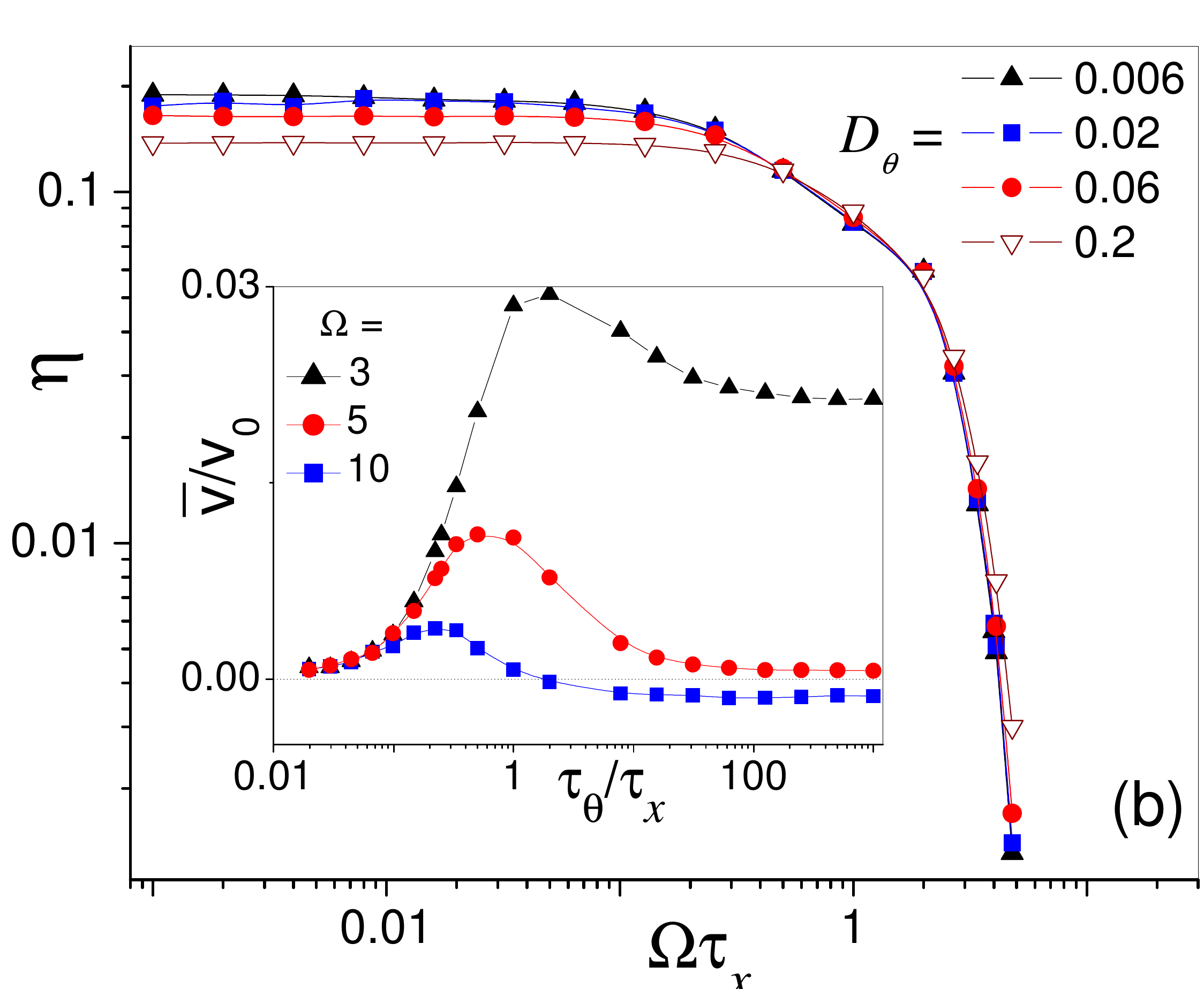}
\caption{(Color online) Rectification of a chiral JP in a triangular
channel: (a) $\eta$ vs. $\tau_{\theta}$ for different $\Omega$; the
same data sets are plotted in a semi-logarithmic (main panel) and
bilogarithmic graph (inset); (b) $\eta$ vs. $\Omega$ for different
$\tau_{\theta}$. Here, $\tau_x\equiv x_L/v_0$, $D_0=0.03$ and the
compartment geometry is as in Fig. \ref{F2}. Inset: $\eta$ vs.
$\tau_{\theta}$ for large $\Omega$, see main panel (b). \label{F5}}
\end{figure}

The LE system of Eq. (\ref{LE}) was first numerically simulated in
for a JP confined to a directed channel made of triangular
compartments with dimensions $x_L \times y_L$ and pore size $\Delta$,
see Fig. \ref{F2}. The compartment aspect ratio was kept constant,
$r=x_L/y_L=1$, and by rescaling the coordinates $x$ and $y$ by an
appropriate factor $\kappa$, $x \to x/\kappa$ and $y\to y/\kappa$,
its dimensions can always be conveniently rescaled to $x_L=y_L=1$ (as
done in Figs. \ref{F2}-\ref{F5} and \ref{F11}). Analogously, by time
rescaling, $t\to v_0 t/\kappa$, one can work with self-propulsion
velocities of constant modulus, $v_0=1$. In conclusion, the output of
our numerical analysis only depends on four characteristic lengths:
the pore width, $\Delta$; the thermal length, $D_0/v_0$; the
self-propulsion length, $l_{\theta}$; and the chiral radius,
$R_\Omega$, all to be compared with the compartment dimensions set to
one. Throughout our simulation work we assumed narrow pores and low
thermal noise, so that the first two lengths play no key role in the
discussion of our results [see Eq. (\ref{furth})]. Equivalently, when
appropriate, instead of the ratios $l_\theta/x_L$ and $R_\Omega/x_L$
one can make use of the dimensionless quantities $\tau_\theta/\tau_x$
and $|\Omega| \tau_\theta$, where $\tau_\theta$ is the
self-propulsion time, $\tau_x=x_L/v_0$ a characteristic compartment
crossing time, and $\Omega$ the chiral angular frequency.

Note that, in view of their orientation, the triangular compartments
of Fig. \ref{F2} tends to funnel the particle to the right (easy-flow
direction) with ${\bar v}>0$. The magnitude of this effect depends on
the modulus and not on the sign of $\Omega$.

{\it Non-chiral Janus particles, $\Omega=0$.} This is the case first
reported in Ref. \cite{MSshort}. When the Janus self-propulsion
length $l_\theta$ is larger than the compartment dimensions, the
particle undergoes several collision, during the time interval
$\tau_\theta$ (Knudsen regime \cite{Brenner}). As the Janus dynamics
gets more sensitive to the compartment asymmetry, the curves ${\bar
v}(\tau_\theta)$ increase monotonously with $\tau_\theta$, until they
level off to an asymptotic upper bound \cite{tumble}, see Fig.
\ref{F3}(a). Most importantly, such asymptotic $\eta$ values are much
larger than the rectification power of the thermal ratchets
investigated in the earlier literature~\cite{RMP2009}.

The impact of thermal noise on the rectification of a JP can be
summarized as follows. In Ref. \cite{MSshort} we showed that for
$l_\theta \ll x_L$ the self-propulsion velocity changes orientation
before the particle slides along a compartment side and through the
pore. Thermal noise, by pulling the particle towards the central lane
of the channel, acts as a lubricant. On the contrary, for $l_\theta
\gg x_L$, thermal fluctuations help the particle overcome the
blocking action of the compartment corners, see Fig. \ref{F2}, thus
suppressing rectification. In the intermediate regime, where $\eta$
is the strongest, these two opposite actions of thermal noise
coexist, as illustrated in Fig. \ref{F3}(b), thus defining an optimal
thermal noise level.

Finally, in view of practical applications, we tested the robustness
of JP rectification in channels with variable degrees of asymmetry:
(i) In Fig. \ref{F4} we modified the compartment geometry by shifting
the corner coordinate, $x_0$, in the range $[0,x_L/2)$ (see inset).
One immediately sees that $\eta$ decreases by only a factor 2 for
$x_0$ up to 0.2; (ii) In our previous report \cite{MSshort} we
studied the consequence of rescaling the $x$ and $y$ compartment
dimensions by a factor $\kappa$. We concluded that rectification is
rather insensitive to $\kappa$ in the Knudsen regime,
$l_\theta>\kappa x_L$. For exceedingly large $\kappa$, the intensity
of the rescaled translational (thermal) noise, $D_0/\kappa v_0$, is
suppressed with respect to the intensity of the rescaled propulsion
noise, $\kappa D_\theta/v_0$, which means that the role of thermal
fluctuation grows negligible on increasing $\kappa$. As a
consequence, in this limit $\eta$ approaches a constant, that is,
${\bar v}$ is inverse proportional to $\kappa$; (iii) In Ref.
\cite{MSshort}, we also reported that for narrow pores, $\Delta \ll
y_L$, $\eta$ slightly decreases with lowering $\Delta$. The
explanation is simple. As the pore shrinks, the compartment sidewalls
grow longer and the particle takes more time to slide along them up
to the exit pore. On the other hand, the negative flow is blocked
mostly at the compartment corners, regardless the actual pore size.
This result indicates that our numerical analysis can be readily
extended to more realistic Janus swimmers of finite radius; (iv) On
running our integration code for $\theta$ randomizing b.c. (not
shown), we obtained substantially smaller $\eta$ values. This is a
consequence of the fact that the persistency of the self-propulsion
velocity is suppressed by the particle collisions against the walls.
This effect gets more pronounced with increasing $D_0$, as thermal
noise causes more wall collisions and, thus, stronger $\theta$
randomization at the boundaries \cite{MSshort}.

{\it Chiral Janus particles, $\Omega>0$.} An angular bias $\Omega$
affects the autonomous ratchet effect discussed so far only in the
strong chirality regime. In Fig. \ref{F5} we present numerical
simulation results for levogyre JP's, $\Omega>0$, diffusing in the
triangular channel of Fig. \ref{F2}. As expected, on increasing
$\tau_\theta$ the rectification power approaches an horizontal
asymptote \cite{tumble}, see Fig. \ref{F5}(a). However, such an
asymptote gets lower at higher $\Omega$, until the curves $\eta=\bar
v/v_0$ and, therefore, $\bar v$ versus $\tau_\theta$ develop a
distinct maximum, see insets of panels (a) and (b). This change marks
the crossover between the regimes of weak and strong chirality.
Indeed, the chiral nature of the JP dynamics can be fully appreciated
when the autocorrelation time of its self-propulsion velocity,
$\tau_\theta/2$, is of the order of the reciprocal of the cyclotron
frequency, $\Omega$, namely for $\tau_\theta \simeq 2/\Omega$. This
simple argument closely locates the maxima of $\bar v$ in both insets
of Fig. \ref{F5}.

The dependence of the rectification power on $\Omega$ is illustrated
in panel (b) of Fig. \ref{F5}. Independently of the level of
translational noise, $D_0$, $\eta$ is largely insensitive to $\Omega$
until to a certain value, after which it suddenly drops to zero. Such
a threshold value, termed here $\Omega_M$, can be estimated by
noticing that on increasing $\Omega$ the chiral radius
$R_\Omega=v_0/|\Omega|$ decreases, until the microswimmer can perform
a full circular orbit inside the compartment, without touching the
channel walls (actually a logarithmic spiral with exponentially small
steps \cite{Lowen}). In the noiseless limit, this happens for
$2R_\Omega \simeq x_L$, that is, $\Omega_M \simeq 2v_0/x_L$. Of
course, this argument holds under the additional condition that
$\Omega_M \tau_\theta>1$, to ensure a sufficiently long
self-propulsion time. This estimate of $\Omega_M$ is in close
agreement with the data of Fig. \ref{F5}(b) and will be used to
explain the peaks developed by $\eta$ in Fig. \ref{F7}(b) of Sec.
\ref{upsidedown}.

Another remarkable result of this section is reported in the inset of
Fig. \ref{F5}(b): At very high $\Omega$, the horizontal asymptote of
$\bar v$ changes sign. This is an instance of the current reversal
phenomenon one often encounters in the ratchet literature
\cite{RMP2009}. When the chiral radius grows so large that the
chirality of the JP does not affect much its pore crossing, $R_\Omega
\gtrsim \Delta$, self-propulsion generates an effective translational
(colored) noise, with time constant $\tau_\theta/2$, which adds to
the white noise of constant $D_0$, as detailed in  Secs.
\ref{diffnonchiral} and \ref{diffchiral}. Under these circumstances,
Eq. (\ref{LE}) thus describes a thermal ratchet. On the contrary, the
rectification mechanism of non-chiral particles described above is
rather reminiscent of a rocked ratchet (this ratchet classification
is reviewed in Ref. \cite{RMP2009}). For a given left-right
asymmetric ratchet potential, rocked and thermal ratchets tend to
generate opposite rectification currents. For this reason the current
reversals depicted in the inset of Fig. \ref{F5} is not totally
unexpected. However, the magnitude of the currents involved is
probably too small to be of practical use.

\begin{figure}[bp]
\centering
\includegraphics[width=0.65\textwidth]{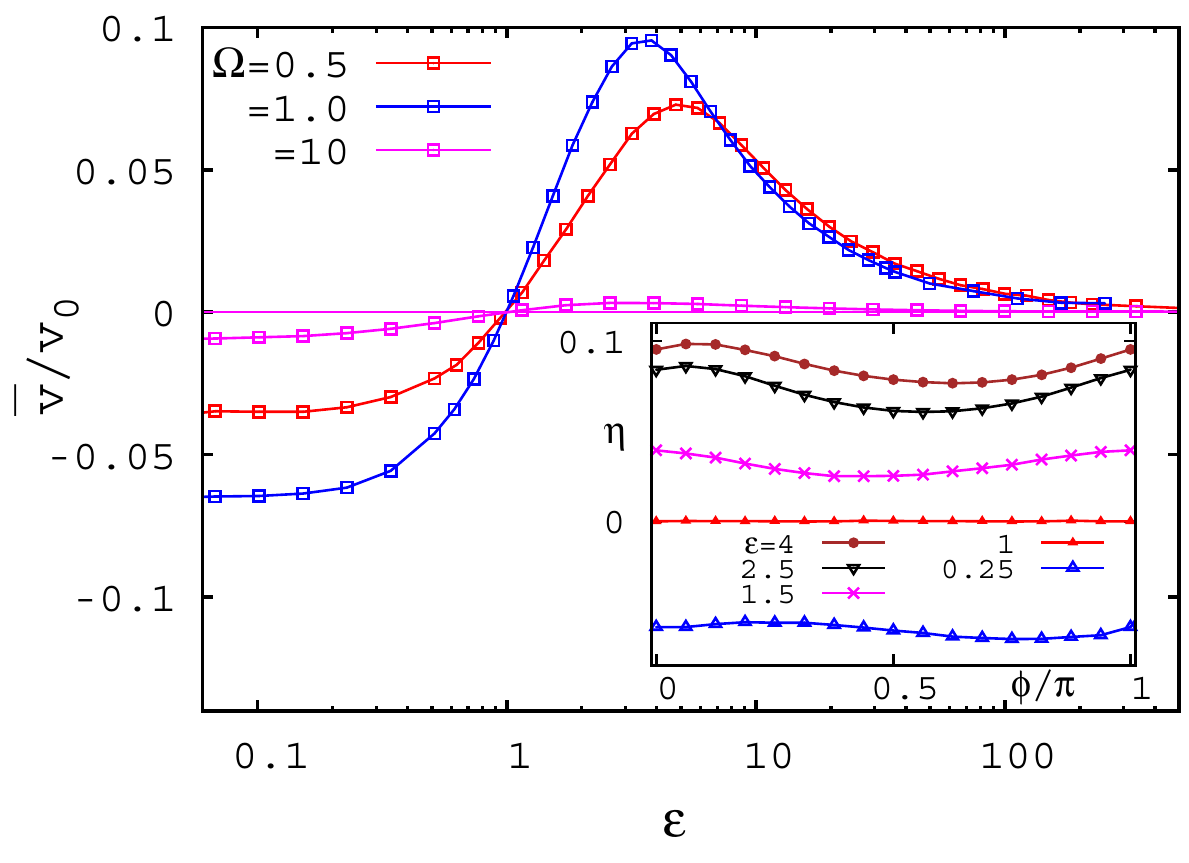}
\caption {(Color online) Rectification of a levogyre JP in the
channel of Eq. (\ref{wx}): $\bar v/v_0$ vs. $\epsilon$ for
$\Delta=0.12$, $D_0=0.01$ and different $\Omega$. We remind that for
a right-left symmetric channel, $\bar v(-\Omega)=-\bar v(\Omega)$.
Other simulation parameters are: $D_\theta=0.3$, $v_0=1$, and
$x_L=y_L=1$. Inset:  $\eta$ vs. $\phi$ for $\Omega=1$ and $\epsilon$
as in the legend. All other simulation parameters are as in the main
panel. \label{F6}}
\end{figure}

\subsection{Upside-down asymmetric channels} \label{upsidedown}

\begin{figure}
\centering
\includegraphics[width=0.8\textwidth]{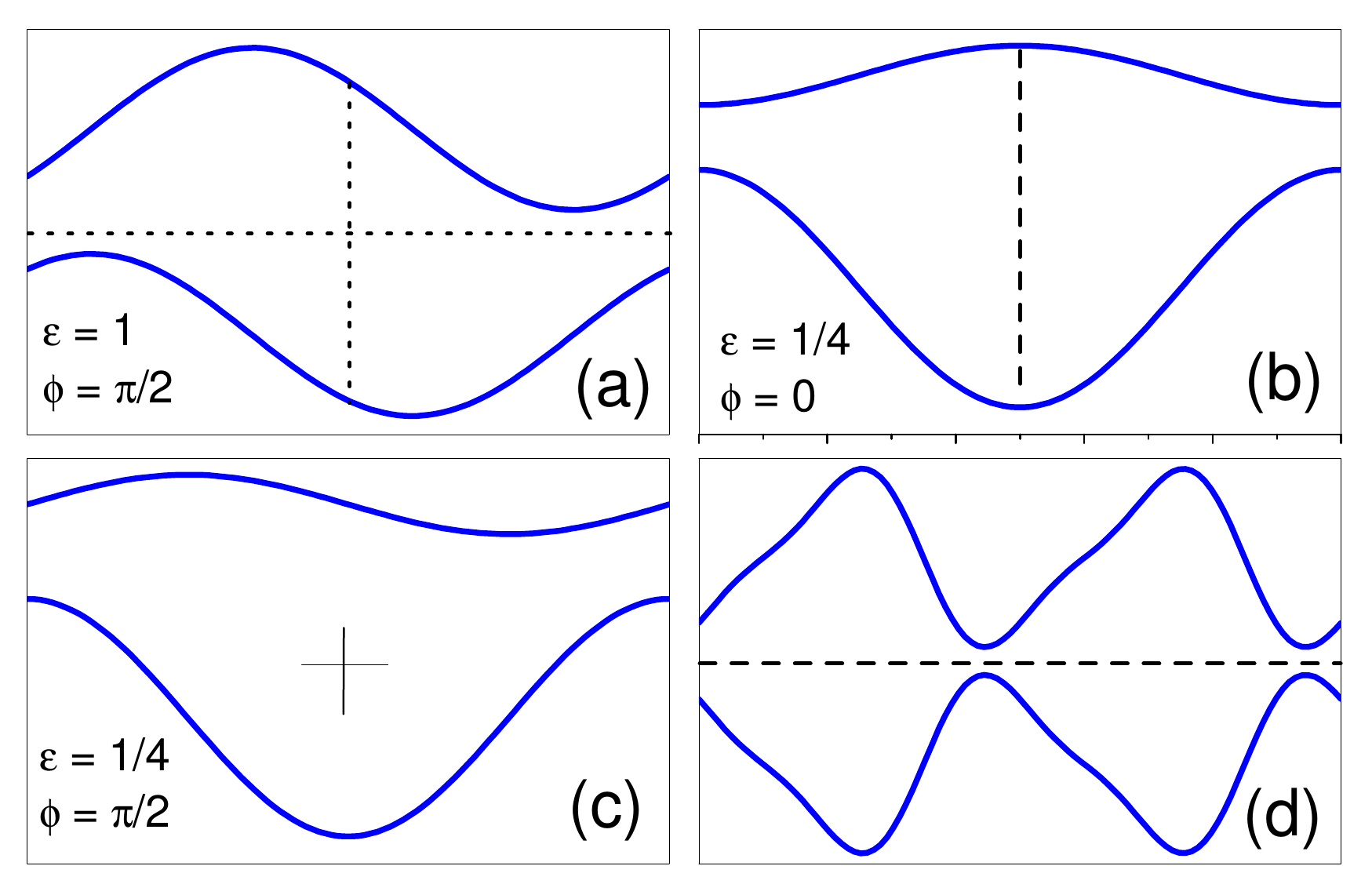}
\caption{(Color online) Compartments of periodically corrugated
channels with different symmetry properties: (a) centro- or
supersymmetric; (b) left-right symmetric; (c) asymmetric; and (d)
upside-down symmetric. The channels walls, $w_{\pm}(x)$, in (a)-(c)
are given by Eqs.~(\ref{wx}) for $\epsilon$ and $\phi$ as reported;
(d) example of upside-down symmetric, left-right asymmetric channel
with $w_{\pm}(x)$ biharmonic sinusoidal functions with components of
period $x_L$ and $x_L/2$.} \label{F8}
\end{figure}

The model Langevin equations (\ref{LE}) have been integrated in Ref.
\cite{SoftMatter} to study the net flow of a levogyre microswimmer
with $\Omega > 0$, confined to the periodic channel of boundaries,
\begin{eqnarray}
\label{wx}
w_+(x) &=& \frac{1}{2} \left [\Delta +\epsilon(y_L-\Delta)\sin^2\left(\frac{\pi}{x_L}x +\frac{\phi}{2} \right )\right ], \nonumber \\
w_-(x) &=& -\frac{1}{2} \left [\Delta
+(y_L-\Delta)\sin^2\left(\frac{\pi}{x_L}x\right) \right ],
\end{eqnarray}
where $x_L$ quantifies the compartment length, $\Delta$ the pore
size, and $y_L$ the channel width. Two additional tunable geometrical
parameters have been introduced in $w_+(x)$, namely, $\phi$ and
$\epsilon$ with $\epsilon \geq 0$, respectively, to shift the
position and tune the amplitude of the upper wall with respect to the
lower one (a few examples of the corresponding channel compartments
are drawn in Fig. \ref{F8}).

When confined to a channel compartment of size smaller than its
self-propulsion length, $l_\theta>x_L$, a chiral microswimmer tends
to align its velocity parallel to the walls
\cite{Bechinger,LowenKumm}, thus generating two boundary flows
oriented to opposite directions, see Fig. \ref{F1}(b). For $\Omega
>0$, the JP is levogyre, which means that the upper and lower boundary
flows are oriented, respectively, to the left and right. The net flow
along the channel axis, $\bar v$, takes the sign of the flow along
the least corrugated boundary, that is $w_+(x)$ for $\epsilon<1$ and
$w_-(x)$ for $\epsilon>1$  and vanishes for $\epsilon=1$. This
mechanism explains the current reversals shown in Fig. \ref{F6}.

\begin{figure}[tp]
\centering
\includegraphics[width=0.49\textwidth]{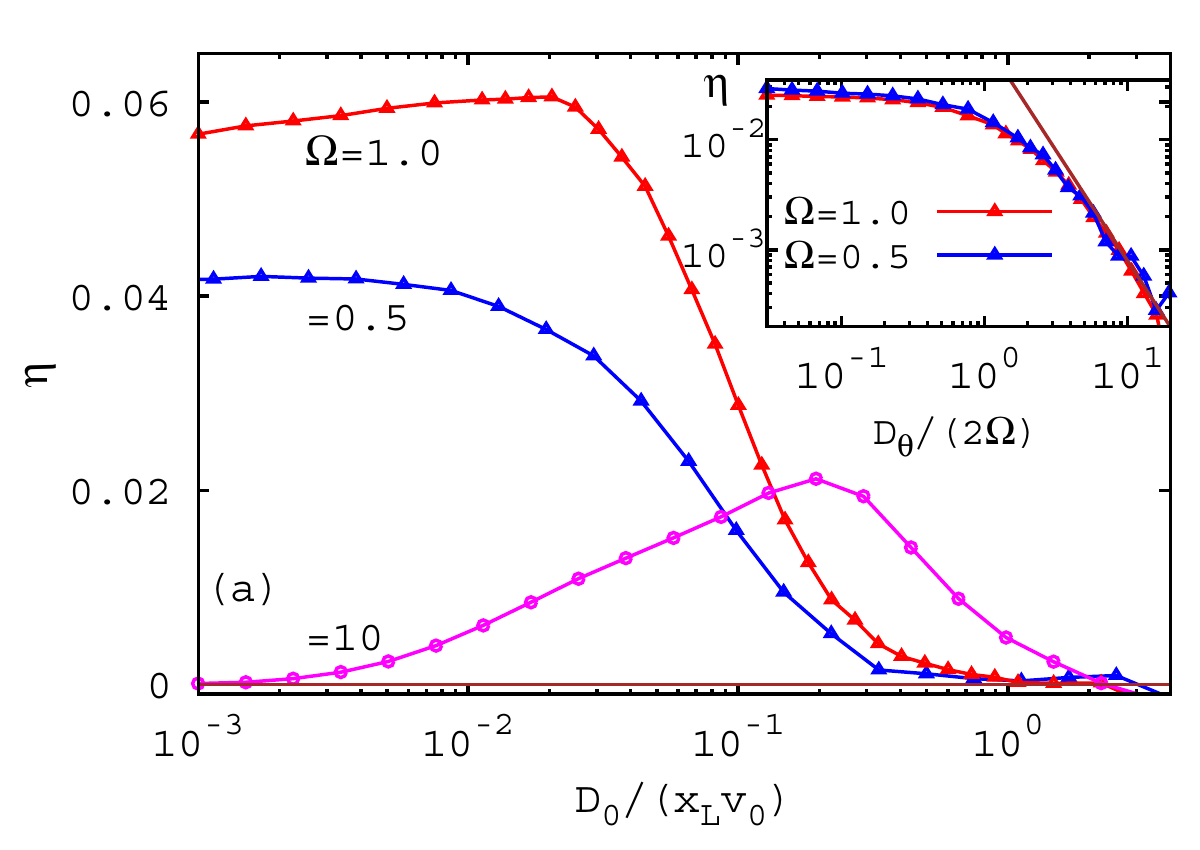}
\includegraphics[width=0.49\textwidth]{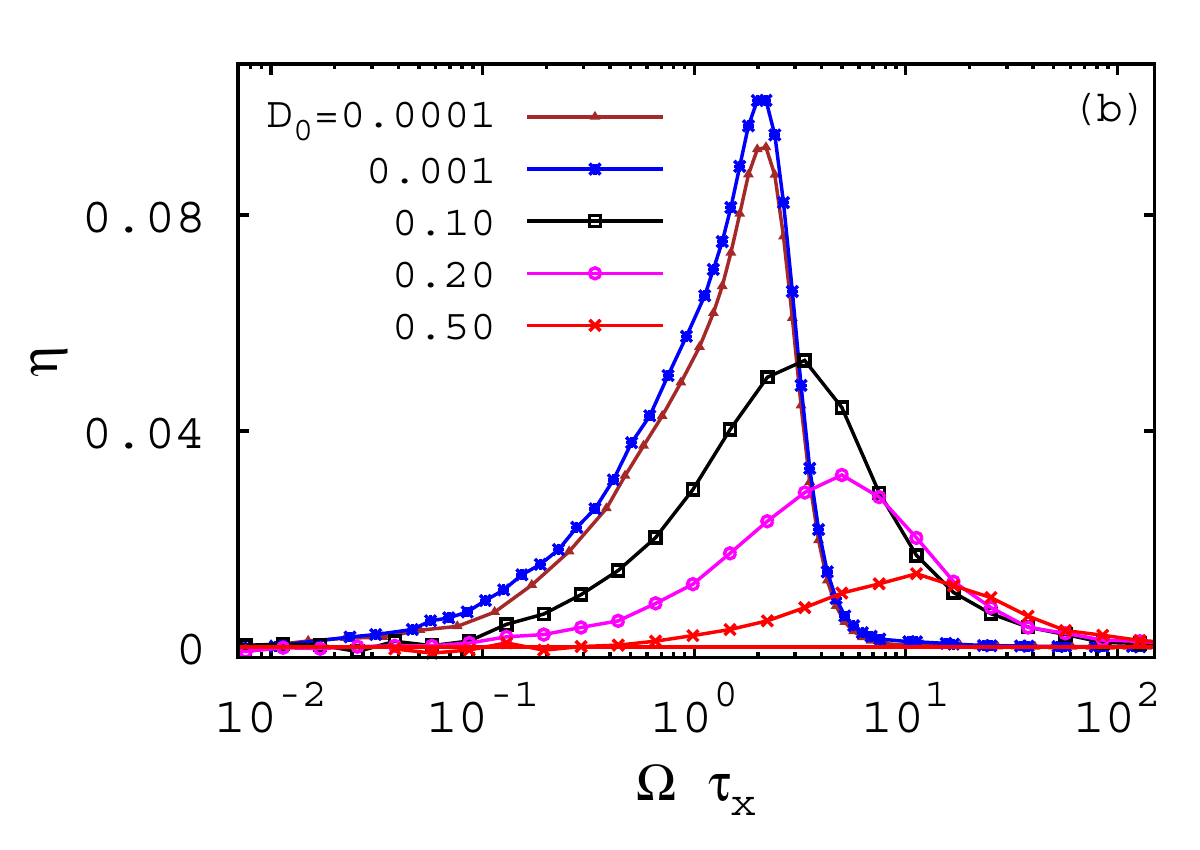}
\caption {(Color online) Optimization of the rectification of a
levogyre JP with $v_0=1$ in the channel of Eq. (\ref{wx}) with
$\epsilon=0.25$, $\phi=0$, and $x_L=y_L=1$: (a) role of noise: $\eta$
vs. $D_0$ for $\Delta = 0.08$, $D_\theta=0.1$, and different $\Omega$
(see legends). Inset: $\eta$ vs. $D_\theta$ for $D_0=0.05$,
$\Delta=0.08$ and different $\Omega$; (b) role of frequency: $\eta$
vs. $\Omega$ for $\Delta=0.12$, $\tau_\theta=10$, $D_0=0.1$,
$\tau_x=x_L/v_0$, and different $D_0$ (see legend). \label{F7}}
\end{figure}

The dependence of  $\eta$ on the fluctuation parameters $D_0$ and
$D_\theta$ is illustrated in Fig. \ref{F7}, adapeted from
\cite{SoftMatter}, and in particular in panel (a). The rectification
power is proportional to $(\Omega \tau_\theta)^{2}$ in the weak
chirality regime, i.e., inverse proportional to $D_\theta^2$ [see
inset of Fig. \ref{F7}(a)]. In the opposite limit of strong
chirality, $\Omega \tau_\theta \gg 1$, $\eta$ approaches a maximum,
which depends on $D_0$, the chiral radius $R_\Omega$ and the
compartment geometry. The curves of $|\bar v|$ versus $D_0$ increase
(decrease monotonically) at high (low) frequency; they all eventually
decay to zero for $D_0 \gg v_0|\Omega|$, no matter what the value of
$\Omega$.

For an optimal value of $\Omega$ and low noise levels, the particle
tends to accumulate against the walls \cite{Fily,MSshort}, with
tangential velocities close to $\pm v_0$. This is the condition of
strong chirality, $|\Omega| \tau_\theta \gg 1$, and low noise,
$v_0|\Omega|/D_0 \gg 1$, where strong autonomous rectification was
first reported \cite{SoftMatter}. Indeed, as the chiral radius
exceeds the compartment dimensions, $R_\Omega \gg x_L$, a strongly
chiral swimmer spends more time drifting between the upper and lower
walls than sliding along them, thus weakening the boundary flows. On
the other hand, when  the chiral radius grows too short, $R_\Omega
\ll x_L$, diffusion occurs mostly away from the boundaries. As a
consequence, in both $R_\Omega$ limits the torque exerted by $\Omega$
becomes ineffective and $\bar v$ tends to vanish. Of course, in the
weak chirality regime, $|\Omega| \tau_\theta \ll 1$, chirality
effects are negligible, altogether.

This behavior is confirmed by the rectification peaks of the curves
$\eta$ versus $\Omega$ displayed in Fig. \ref{F7}(b). On decreasing
$D_0$ the $\eta$ peak shifts toward a limiting value, where it is the
most pronounced. This is the $\Omega$ threshold  value, $\Omega_M$,
introduced in the previous Sec. \ref{leftright}. Indeed, for $\Omega
\simeq \Omega_M$ we know already that the microswimmer can perform a
circular orbit, without being captured by the boundary layers.
Moreover, by closing its orbit inside a compartment, the swimmer gets
trapped there, which explains the sudden drop of $\eta$ for $\Omega
\geq \Omega_M$. The peak in the curves $\bar v$ versus $\epsilon$
for $\epsilon>1$ and constant $\Omega$ reported in Fig. \ref{F6} can
be interpreted in the same way \cite{SoftMatter}.

We have already seen that thermal noise disrupts the boundary flows
by kicking the particle inside the compartment. Moreover, it also
perturbs its circular orbits by making them spiral faster and their
centers diffuse. That is why, on increasing $D_0$, the $\Omega$-peak
tends to shift to higher $\Omega$ (i.e., smaller $R_\Omega$) and
diminish in height, as shown in Fig. \ref{F7}(b) and anticipated in
Fig. \ref{F7}(a).

\subsection{Channel asymmetry requirements} \label{asymmetry}

In our interpretation the rectification process is governed by the
boundary flows, and therefore by the spatial symmetry of the channel
walls \cite{Denisov}. This picture is consistent with rigorous
symmetry arguments. First of all we notice that the 2D channel
compartments in Fig. \ref{F8} can be asymmetric under inversion of
either the $y$ axis ($y\to -y$, upside-down asymmetric), panel (b),
of the $x$ axis ($x\to -x$, right-left asymmetric), panel (d), or
both, panel (a,c). Most remarkably, compartment (a), while both
upside-down and right-left asymmetric, is invariant under any pair of
$x$ and $y$ axis inversions, namely, it is centro-symmetric. On
combining the symmetry properties of the model dynamics, Eq.
(\ref{LE}), with those of the channel compartment, we arrive at a few
interesting conclusions:

(i) With reference to the right-left symmetric compartment (b), we
notice that Eqs. (\ref{LE}) are invariant under the transformations
$x\to -x$ and $\theta \to \pi - \theta$ or, equivalently, $\Omega \to
-\Omega$, which leave the channel also invariant; hence $\bar
v(-\Omega)= -\bar v (\Omega)$. As a consequence, nonchiral JP's
cannot be rectified in compartment (b), as, clearly,  $\bar v(0)=0$.
This last property holds in a wider sense, as discussed in item (v)
below.

(ii) Analogously, for an upside-down symmetric channel one concludes
that $\bar v(\Omega)= \bar v (-\Omega)$. The consequence of this last
symmetry relation is that for the triangular channel of Sec.
\ref{leftright} $\bar v$ can only be a function of $\Omega^2$, which
explains the flat branch of the $\eta$ curves with $\Omega <
\Omega_M$, plotted in Fig. \ref{F5}(b).

(iii) By shifting the channel walls $w_\pm(x)$ in Eq. (\ref{wx}) by a
length $\phi$, one can easily prove the additional symmetry relations
$\bar v(\phi, \Omega)=\bar v(-\phi,\Omega)$ for right-left symmetric
compartments [compare compartments (b) and (c) in Fig. \ref{F8}] and
$\bar v(\phi, \Omega)=\bar v(-\phi,-\Omega)$ for upside-down
symmetric compartments. As displayed in the inset of Fig. \ref{F6},
$\bar v$ is weakly modulated by a relative shift of the walls,
$\phi$, and so are the boundary flows.

(iv) For a centro-symmetric compartment, (a), both parity relations
hold simultaneously; hence, $\bar v(\Omega)=0$. Numerical simulations
for the channel of Eq. (\ref{wx}) with $\epsilon=1$ and any $\phi$
support this conclusion.

(v) From items (i) and (ii) one is led to conclude that $\Omega \neq
0$ is a necessary condition for JP rectification in the right-left
symmetric channels but not in the upside-down symmetric ones. We
observed that this condition applies, in fact, to a wider class of
compartments, which includes compartments (a)-(c) of Fig. \ref{F8}.
For the sake of an argument, we assume first that self-propulsion is
switched off, i.e., $v_0=0$. As apparent from Fick-Jacobs reduction
technique \cite{ChemPhysChem,Fick,Jacobs,Zwanzig}, diffusion along a
smooth directed channel depends on the modulating function
$\sigma(x)=w_+(x)-w_-(x)$. If $w_{\pm}(x)$ are sinusoidal functions
of period $x_L$, so is their difference, $\sigma(x)$. Accordingly,
the reduced longitudinal particle dynamics would be mirror symmetric.
As self-propulsion is switched on, the diffusing particle is subject
to an additional time-correlated noise, which does break the time
symmetry. However, due to the mirror symmetry of the reduced particle
dynamics along the channel axis, this does not suffice to ensure
rectification of a JP with $\Omega=0$. Put differently, a breach of
the right-left symmetry of the channel compartment, does not suffice
to rectify nonchiral JP's.

\section{Diffusion of active microswimmers} \label{diffusion}

As a measure of the efficiency of the autonomous rectification
mechanism we now analyze the dispersion of a JP along the channel
axis~\cite{Machura}. This is an important issue experimentalists
address when trying to demonstrate rectification: Indeed, drift
currents, no matter how weak, can be detected over an affordable
observation time only if the relevant dispersion is sufficiently
small. To this purpose we compute the transport diffusivity of a JP
in a channel defined as
\begin{equation} \label{diffeff}
D_{\rm ch}=\lim_{t\to \infty}[\langle x^2(t)\rangle - \langle x(t)
\rangle^2]/(2t).
\end{equation}

\subsection{Diffusion of nonchiral microswimmers, $\Omega=0$} \label{diffnonchiral}

\begin{figure}[tp]
\centering
\includegraphics[width=0.60\textwidth]{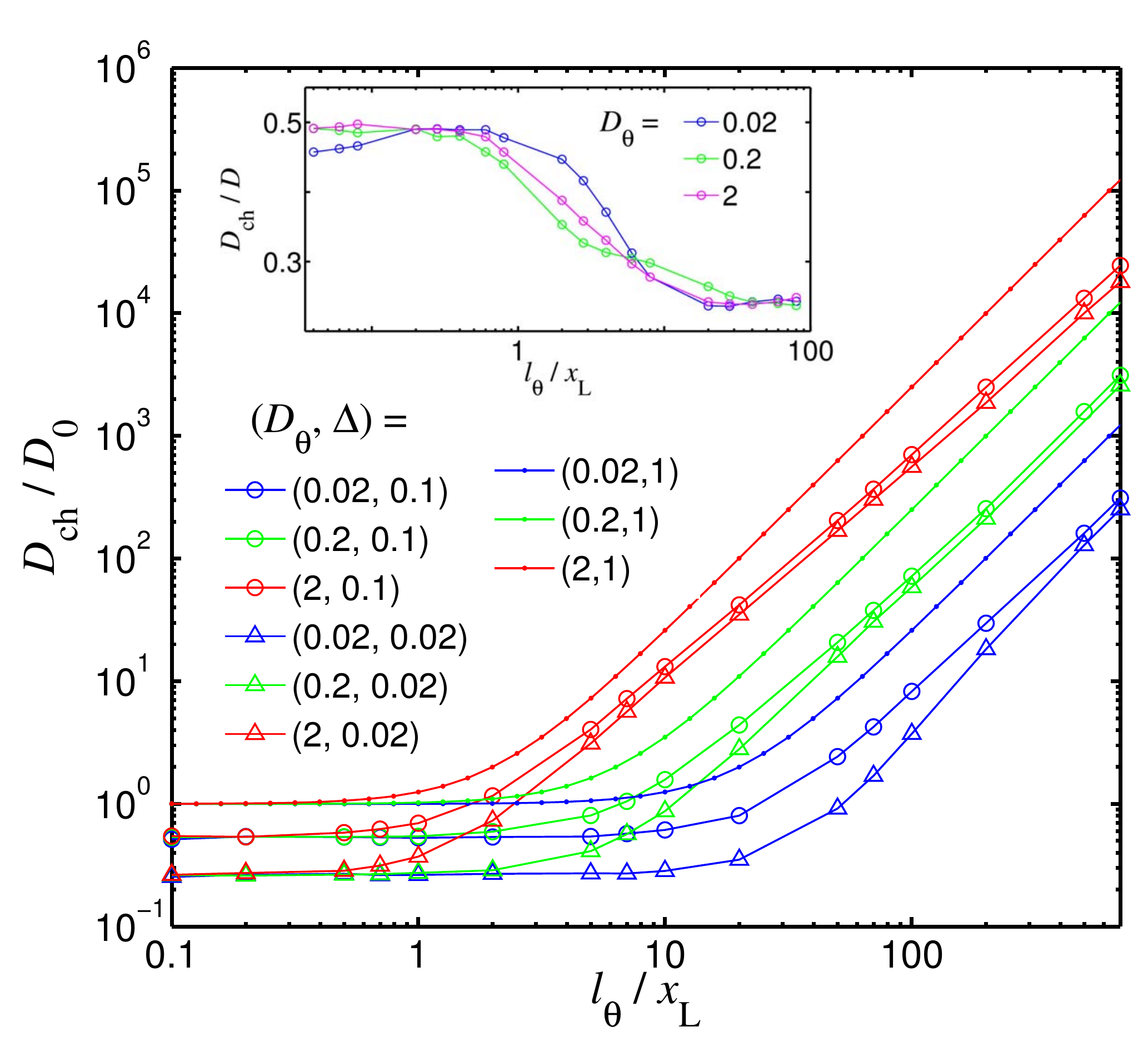}
\caption{(Color online) Diffusion of a levogyre JP in a fully
symmetric sinusoidal channel with $w_\pm(x)$ given in Eq. (\ref{wx}):
$D_{\rm ch}$ vs. $l_\theta$ for different $D_\theta$ and $\Delta$.
Note that $\Delta=1$ represents the limiting case of a straight
channel of width $\Delta=y_L=1$ and here $D_{\rm ch}=D$, see Eq.
(\ref{diff1}). Other simulation parameters are $x_L=y_L=1$,
$\epsilon=1$, and $D_0=0.01$. Inset: $D_{\rm ch}/D$ vs $l_\theta/x_L$
for different $D_\theta$ (see legend) and $D$ defined in Eq.
(\ref{diff1}). All remaining simulation parameters are as in Fig.
\ref{F10}, with the numerical estimates $\kappa_0=0.48$ and
$\kappa_s=0.23$. We checked that our data do better approach the
predicted power law, $D_{\rm ch} \propto l_{\theta}^2$, on further
increasing $l_\theta$. \label{F9}}
\end{figure}

A full analytical investigation of the model of Eq. (\ref{LE}) is out
of question even in the bulk and for nonchiral JP's, $\Omega=0$.
However, on noticing that \cite{MSshort,Marchetti} $$\langle \cos
\theta (t) \cos \theta (0) \rangle=\langle \sin \theta (t) \sin
\theta (0) \rangle=(1/2)e^{-|t|D_\theta},$$ it follows immediately
that the self-propulsion velocity components $v_{0,x}=v_0 \cos
\theta$ and $v_{0,y}=v_0 \sin \theta$ can be regarded as the
components of a 2D non-Gaussian noise ${\vec \xi_{s}}(t)$ with zero
mean, $\langle \xi_{s,i}(t)\rangle=0$, and finite-time correlation
functions, $\langle
\xi_{s,i}(t)\xi_{s,j}(0)\rangle=2(D_s/\tau_\theta)\delta_{ij}e^{-2|t|/\tau_\theta}$,
where $D_s=v_0^2\tau_\theta/4$ and $\tau_\theta=2/D_\theta$. In the
bulk the first two LE of Eq. (\ref{LE}) are statistically independent
and, therefore, a nonchiral particle diffuses according to F\"urth's
law
\begin{equation}
\label{furth} \langle \Delta \vec{r}(t)^2\rangle = 4
(D_0+v_0^2\tau_\theta/4)t
+(v_0^2\tau_\theta^2/2)(e^{-2t/\tau_\theta}-1),
\end{equation}
with $\Delta \vec{r}(t) \equiv \vec{r}(t)-\vec{r}(0)$. Accordingly,
the approximate equality $\langle \Delta \vec{r}(t)^2\rangle = 4Dt$,
holding for $t \gg \tau_\theta$, defines the particle bulk
diffusivity,
\begin{equation}
\label{diff1} D=D_0+D_s\equiv D_0+{v_0^2\tau_\theta}/{4}.
\end{equation}
Of course, if the JP diffuses in a non-corrugated channel, say, with
$w_+(x)=w_-(x)=y_L/2$, then $D_{\rm ch}=D$, as confirmed by the
simulation data of Fig. \ref{F9} (dashed curves).

\begin{figure}[tp]
\centering
\includegraphics[width=0.60\textwidth]{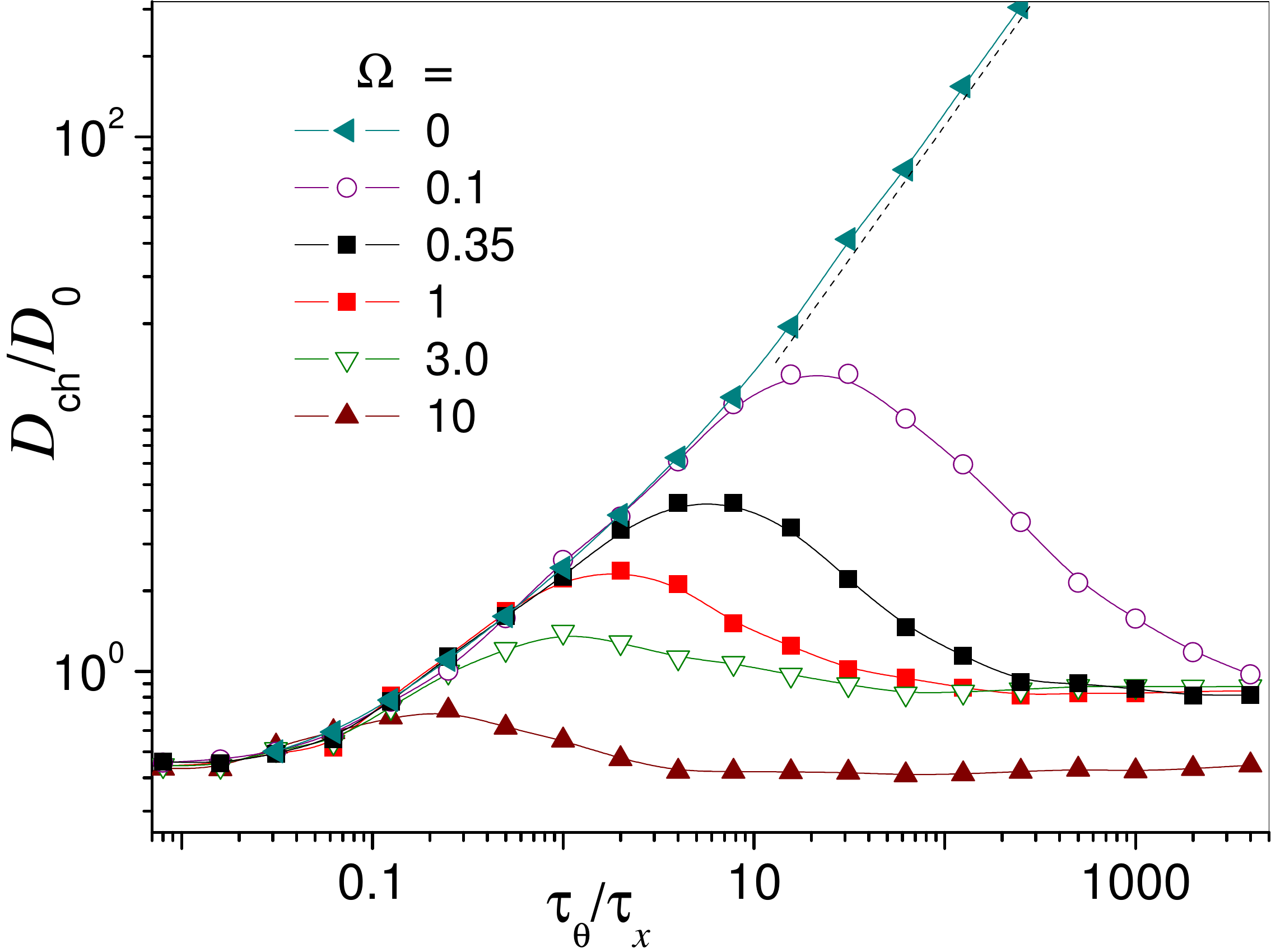}
\caption{(Color online) Diffusion of a levogyre JP in the sinusoidal
channel of Eq. (\ref{wx}): $D_{\rm ch}/D_0$ vs. $\tau_\theta$ for
different $\Omega$. Here, $\tau_x\equiv x_L/v_0$, $v_0=1$,
$D_0=0.05$, $\Delta=0.08$ and $\epsilon=1$. The dashed line
represents the asymptotic linear power-law of Eq.(\ref{diff1}).}
\label{F10}
\end{figure}

When confined to a sinusoidal channel, the particle diffusivity is
suppressed by the geometric constrictions represented by the pores,
see Figs. \ref{F9} for nonchiral and \ref{F10} for chiral JP's. In
the absence of self-propulsion, $v_0=0$, the bulk diffusivity of a
{\it non-chiral} JP, Eq. (\ref{diff1}), is $D=D_0$ and the channel
diffusivity can be written as $D_{\rm ch}=\kappa_0 D_0$, with
$\kappa_0$ a well studied function of $\Delta$ and $D_0$
\cite{Schmid,Bosi}. In the opposite limit, $v_0 \to \infty$, the
process is governed by self-diffusion, that is, $D\simeq D_s$ and,
accordingly, $D_{\rm ch}=\kappa_s D_s$, with $\kappa_s$ much less
sensitive to the pore constriction than $\kappa_0$. Both asymptotic
regimes of $D_{\rm ch}$ are illustrated in Fig. \ref{F9} for
different values of $\Delta$ and $D_\theta$.

\begin{figure}[bp]
\centering
\includegraphics[width=0.6\textwidth]{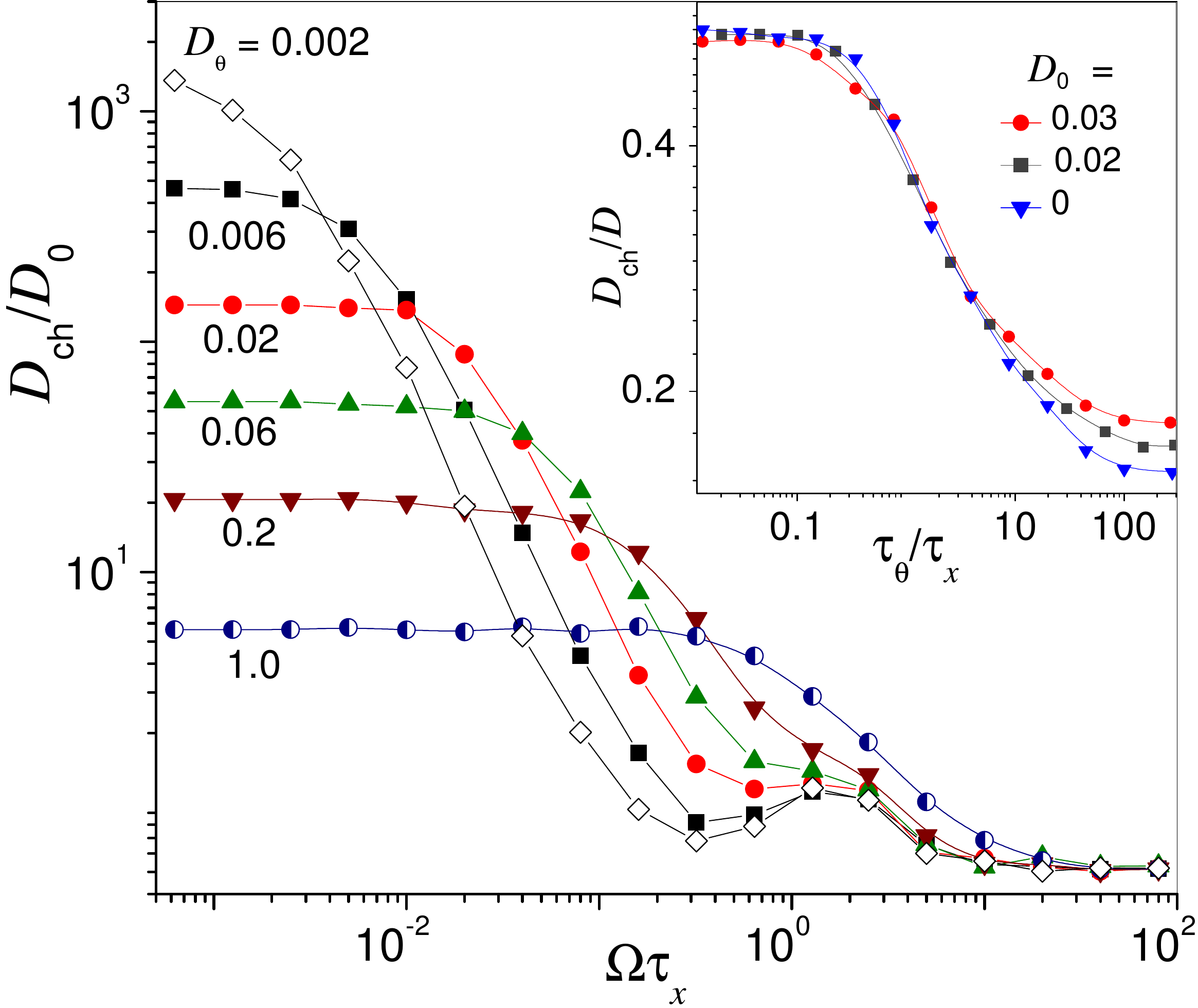}
\caption{(Color online) Diffusion of a levogyre JP in the triangular
channel of Fig. \ref{F2}: $D_{\rm ch}/D_0$ vs. $\Omega$ for different
$D_\theta$. Here, $\tau_x\equiv x_L/v_0$, $D_0=0.03$ and the
remaining simulation parameters are as in Fig. \ref{F2}.  The
diffusion of a nonchiral JP, $\Omega=0$, is shown in the inset for
different $D_0$: $D_{\rm ch}/D$ vs. $\tau_\theta$ with $D$ defined in
Eq. (\ref{diff1}).} \label{F11}
\end{figure}

This picture does not depend much on the actual compartment geometry.
This conclusion is corroborated by our simulations for different
channel geometries. The channel diffusivity of a nonchiral JP in the
triangular channel of Sec. \ref{leftright} is plotted in the inset of
Fig. \ref{F11}. Here, too, in the limit of short self-propagation
times, $\tau_\theta \ll \tau_x$, $D$ approaches $D_0$, while we know
that $\eta$ drops to zero, see Fig. \ref{F3}(a). This regime amounts
to an ordinary unbiased Brownian motion occurring in a triangular
channel, for which $D_{\rm ch}/D_0 = \kappa_0$, with $\kappa_0$ a
function of the compartment geometry (for the compartment of Fig.
\ref{F2}, $\kappa_s \simeq 0.55$ \cite{Savelev}). In the opposite
limit, $l_\theta \gg x_L$, as $\eta$ approaches its horizontal
asymptote of Fig. \ref{F3}(a), $D$ grows like $D_s$, but the ratio
$D_{\rm ch}/D$ decreases toward a ($D_0$-dependent) lower bound,
$\kappa_s\simeq 0.16$.

We give next a simple phenomenological argument to compare the
large-$l_\theta$ behaviors of $\eta$ and $D_{\rm ch}/D_0$ in the
triangular channel of Fig. \ref{F2}. The extension of this argument
to the case of the sinusoidal channel considered in Figs. \ref{F9}
and \ref{F10} is given in Ref.\cite{thesis}. Consistently with the
estimate of the bulk active diffusion, $D_s=v_0^2\tau_\theta/4$, we
can assume that a channeled JP self-propels itself to the right and
left, alternately, with time constant $\tau_\theta/2$. When confined
to a channel compartment, its effective self-propulsion velocities to
the right/left are, respectively, $v_{R,L}=\mu_{R,L}v_0$ with
mobility constants $\mu_{R,L}$, which depend on the compartment
geometry. In terms of the right/left mobility, the rectification
power of Eq.(\ref{RP}) reads $\eta=(\mu_{R}-\mu_{L})/2$, and the
corresponding channel diffusivity \cite{Borromeo1,Borromeo2} $D_{\rm
ch}/D_0=(\mu_{R}+\mu_{L})^2/2$. In the absence of thermal noise,
$D_0=0$, for the triangular compartment of Fig. \ref{F2} one can make
use of the approximations $\mu_L=0$  and $\mu_R=\cos^2
\alpha/\sqrt{2}$ \cite{thesis}. The ensuing estimates for $\eta$ and
$D_{\rm ch}$ at zero noise, $\eta^{(0)}=0.28$ and $\kappa_s=D_{\rm
ch}/D_0=0.16$, reproduce fairly closely the relevant asymptotes of
Figs. \ref{F3} and \ref{F11}. It should be remarked that, as
discussed in Sec. \ref{leftright}, for $D_0>0$ thermal noise tends to
suppress $\eta$, $\eta<\eta^{(0)}$, but increase $D_{\rm ch}$,
$D_{\rm ch}>D_{\rm ch}^{(0)}$.

\subsection{Diffusion of chiral microswimmers, $\Omega \neq 0$}
\label{diffchiral}

The transport diffusivity of {\it chiral} JP's is illustrated in
Figs. \ref{F10} and \ref{F11}. The dependence of $D_{\rm ch}$ on
$\Omega$ well summarizes the different chiral regimes discussed in
Sec. \ref{rectificaton}. We pointed out that chiral effects are
observable only if the self-propulsion time constant $\tau_\theta$ is
long enough, that is $|\Omega|\tau_\theta \gg 1$ or $R_\Omega \ll
l_\theta$. On the other hand, when the chiral radius $R_\Omega$ grows
smaller than the compartment dimensions, $R_\Omega \ll x_L$, or
$|\Omega|\tau_x \gg 1$, chirality suppresses active transport. In
general, chiral self-propulsion effects are appreciable for
$\tau_\theta \gg \tau_x$ \cite{MSshort}.

In addition, we remark here that the $\Omega$-dependence of the bulk
diffusivity of a chiral particle can be obtained from Eq.
(\ref{furth}) by applying to our model the approach of Refs.
\cite{Taylor,Kur}, namely
\begin{equation}
\label{diff2}
D(\Omega)=D_0+\frac{v_0^2/\tau_\theta}{(2/\tau_\theta)^2+\Omega^2},
\end{equation}
where $D(0)$ coincides with $D$ in Eq. (\ref{diff1}) \cite{thesis}.

Similarly to the case of the nonchiral JP's discussed in the previous
section, the confining action of the channel corrugations tends to
suppress the particle diffusivity in the channel, $D_{\rm
ch}(\Omega)$. Here too the symmetry of the walls plays no major role.
For this reason we analyze the effects of chirality on active
diffusion by discussing simulation data for a sinusoidal (Fig.
\ref{F10}) and triangular channel (Fig. \ref{F11}) on the same foot.

With these premises the main features of the curves $D_{\rm ch}$
versus $\tau_\theta$ in a sinusoidal channel at constant $\Omega$,
Fig. \ref{F10}, are readily explained: (i) The curve $\Omega = 0$
reproduces the situation of Fig. \ref{F9} with $D_{\rm ch}$ growing
linearly with $\tau_\theta$; (ii) For $D_0 \ll D_s$ the channel
diffusivity is proportional to $D(\Omega)$, that is $D_{\rm
ch}(\Omega) = \kappa_s D(\Omega)$, like for the nonchiral JP's; (iii)
Moreover, the corresponding maxima occur for $|\Omega| \tau_\theta =
2$, see Eq. (\ref{diff2}), with $D_{\rm ch}^{\rm
max}=D(2/\tau_\theta)\simeq \kappa_s D_s/2$; (iv) For finite $\Omega$
the activated diffusivity in the channel is suppressed both for
$\tau_\theta \to 0$ and $\tau_\theta \to \infty$. Accordingly,
$D_{\rm ch} (\Omega) \to \kappa_0 D_0$, where $\kappa_0$ has been
defined in Sec. \ref{diffnonchiral}.

The $\Omega$ dependence of $D_{\rm ch}$ in the triangular channel of
Fig. \ref{F2} at constant $\tau_\theta$ is also consistent with the
above interpretation of chirality effects on channel diffusivity. In
addition, the data sets of Fig. \ref{F11}  show that: (v) In the
regime of weak chirality, $|\Omega|\tau_\theta \ll 1$, the $\Omega$
dependence of $D_{\rm ch}$ is negligible, as it was for $\eta$ in
Fig. \ref{F5}(b). $D_{\rm ch}$ starts decreasing appreciably only for
$|\Omega|\tau_\theta \gtrsim 2$, in coincidence with the $\eta$
maxima  displayed in the insets of Fig. \ref{F5}; (vi) In the regime
of strong chirality, $|\Omega|\tau_\theta \gg 1$, $D(\Omega) \to
D_0$, so that $D_{\rm ch} \simeq \kappa D_0$ with $\kappa \simeq
0.55$ (see discussion of Fig. \ref{F9}); (vii) Finally, small
diffusivity peaks emerge also for $|\Omega|\tau_\theta \gg 1$. They
are centered around $\Omega_M$ and correspond to the sudden drop in
the rectification power [Fig. \ref{F5}(b), main panel] that occurs
when $R_\Omega$ grows shorter than $x_L$.

\section{Conclusions} \label{conclusions}

We numerically simulated the transport of artificial active
microswimmers diffusing along a narrow periodically corrugated
channel. Key transport quantifiers, like rectification power and
diffusivity, strongly depend on the particle self-propulsion
mechanism and the channel compartment geometry. Applications of such
control technique are within the reach of today's technology.
Specialized microfluidic circuits can be designed, for instance, to
guide chiral microswimmers to a designated target. The same technique
can be utilized to fabricate monodisperse chiral microswimmers
(presently a challenging technological task). By the same token,
microswimmers capable of inverting chirality upon binding to a load,
can operate as chiral shuttles along a suitably corrugated channel
even in the absence of gradients of any kind.

The model analyzed here should be regarded as a stepping stone for
more challenging generalizations and sophisticated comparisons with
ongoing experimental work. Among the issues one should address next
we mention: (i) {\it diffusion gradients.} Either the channel profile
or the local inhomogeneities responsible for self-propulsion can be
graded so as to generate an $x$-dependent channel transport diffusion
coefficient, $D_{\rm ch}(x)$, which adds to the ratchet effect
discussed in Sec. \ref{rectificaton}; (ii) {\it hydrodynamic
effects.} We ignored the role of the suspension fluid flowing around
the moving microswimmer. An accurate account of microfluidic effects
is likely to selectively impact the particle boundary flows along a
corrugated channel wall as well as the translocation of finite size
JP's through a narrow pore; (iii) {\it wall interactions.} The
sliding b.c. implemented in our simulation code are known to
reproduce rather closely certain experimental conditions, but surely
are not granted in all setups under investigation. Particle
translocation through narrow constrictions may be extremely sensitive
to the particle-wall interactions, which thus affect both active
rectification and diffusion in corrugated channels.

\section*{Acknowledgements}
X.A. has been supported by the grant Equal Opportunity for Women in
Research and Teaching of the Augsburg University. P.H. and G.S.
acknowledge support from the cluster of excellence Nanosystems
Initiative Munich (NIM). Y.L. was supported by the NSF China under
grants No. 11347216 and 11334007, and by Tongji University under
grant No. 2013KJ025. F.M. thanks the Alexander von Humboldt Stiftung
for a Research Award. All authors thank Riken's RICC for
computational resources.

\end{document}